\documentclass[aps,twocolumn,showpacs,showkeys,superscriptaddress]{revtex4-1}  

\usepackage{epsfig}

\begin{document} 
   
\title{Efficient simulated tempering with approximated weights: Applications
to first-order phase transitions}  
  
\author{A. Valentim} 
\affiliation{Departamento de F\'{\i}sica, Universidade Federal do Paran\'a, CP 19044, 81531-980 Curitiba-PR, Brazil} 
\author{Claudio J. DaSilva}
\affiliation{Instituto Federal de Educa\c{c}\~ao, Ci\^encia e Tecnologia de Goi\'as, C.P. 74130-012, Goi\^ania, Goi\'{a}s, Brazil} 
\author{Carlos E. Fiore} 
\email{fiore@if.usp.br}             
\affiliation{Instituto de F\'{\i}sica, Universidade de S\~{a}o Paulo, C. P. 66318 05315-970 S\~{a}o Paulo, S\~{a}o Paulo, Brazil}   

\date{\today}   
   
\begin{abstract}  
Simulated tempering (ST) has attracted a great deal of attention in the last years, due to its capability to allow systems with complex dynamics to escape from regions separated by large entropic barriers.  However its performance is strongly dependent on basic ingredients, such as the choice of the set of temperatures and their associated weights. Since the weight evaluations   are not trivial tasks, an alternative approximated approach was proposed by Park and Pande (Phys. Rev. E {\bf 76}, 016703 (2007)) to circumvent this difficulty. Here we present a detailed study about this procedure by comparing  its performance with 
exact (free-energy) weights and other methods, its dependence on the total replica number $R$ and on  the temperature set. The ideas above are analyzed in four distinct lattice models presenting strong first-order phase transitions, hence constituting ideal examples in which the performance of algorithm is fundamental. In all cases, our results reveal that approximated weights work properly in the regime of larger $R$'s. On the other hand, for sufficiently small $R$ its performance is reduced and the systems do not cross
properly the free-energy barriers.  Finally, for estimating reliable temperature sets,
we consider a simple protocol proposed at Comp. Phys. Comm.
{\bf 128}, 2046 (2014).
\end{abstract}  
 
 
  
\maketitle   
  
\section{Introduction}

Although Monte Carlo method has become  probably the most common tool for studying phase transitions and critical phenomena, in practice its usage is not so simple,  
whenever standard algorithms (e.g. Metropolis) are used.
Despite the simplicity and generality, they lead to difficulties close to the emergence of phase transitions.
For instance, alternative procedures are typically required, specially in the case of systems 
 with microscopic  configurations  separated by valleys and hills in the free-energy
landscape \cite{binder,metr,spinglass,proteins}. 
Cluster algorithms \cite{sw,bouabci}, 
multicanonical \cite{berg}, Wang-Landau \cite{wang} and tempering methods  are some examples of proposals to overcome these difficulties. In particular,  parallel tempering (PT) \cite{nemoto} and simulated tempering (ST) \cite{parisi} enhanced sampling methods have drawn attention  due to their generality and  simplicity when compared with the previous examples. Their basic idea consists of using configurations from high temperatures for systems at low temperatures, allowing in principle the dynamics to escape from metastable states and providing an appropriate visit of the 
configuration space.
 In particular, distinct aspects of tempering methods have been explored in the last years, aiming at better understanding of efficiency and pertinence. For instance, the role of temperature sets for the PT case was investigated in Refs. \cite{kone, helmut,sabo,fiorejcp}, whilst the importance of non-adjacent exchanges was taken into account in Refs. \cite{juan,juan1,calvo,fiore8,fiorejcp}. In addition the efficiency and comparison between tempering methods were  considered in Refs. \cite{maJCF,fiore10,rosta}.

Focusing our attention in the ST we face one of its main difficulties, namely the evaluation of the free-energy weights,  required for an uniform sampling to all temperatures. Despite the development of alternative techniques, their applicability for more complex systems still poses a hardship. In some cases \cite{ma,maJCF}, the accumulation of histograms (of a given quantity) and previous simulations are necessary to calculate (or to estimate) the input parameters that guarantee a sufficient number of visits to all temperatures. In such cases, the  weights are set arbitrarily but a knowledge of the partition function $Z_i$ 
at each temperature is required and a flattening histogram  based on a random walk in the parameter (temperature or energy) space is used to obtain a satisfactory estimation of $Z_i$.  In Refs. \cite{fiore11,sauerwein}, the partition function is exactly  valued through numerical simulations, taking into account its relationship with the largest eigenvalue $\lambda^{(0)}$  of the transfer matrix ${\mathcal T}$. Although the evaluation of $\lambda^{(0)}$ is possible for lattice-gas systems, its extension for more complex cases (e. g. off-lattice systems) is not straightforward. In contrast to previous ``exact'' approaches, Park and Pande \cite{pande} proposed an approximated tool of estimating weights, based on the average system energy. Since the mean energy is easily obtained for any system (including lattice and off-lattice models),  it constitutes a considerable simplification over the free-energy  case.
 Nevertheless, there are some fundamental points that need to be understood in order to make it a promptly useful method. The first one is how this procedure compares itself with 
using free-energy weights? The second one is under what conditions does it provide equivalent results to those obtained from free-energy weights? An additional point is if it is possible to obtain proper temperature set that yields precise results under lower computational cost.
 To answer the aforementioned points, we have analyzed, under the ST with approximated weights, four distinct lattice models, namely, Blume-Capel (BC) and Blume-Emery-Griffiths (BEG) \cite{BEGMODEL}, Bell-Lavis (BL) \cite{bell,fiore-m} and associating lattice  gas (ALG) water models \cite{alg,algs}. The former two are interesting tests, due to the existence of very precise results available from cluster algorithms \cite{cluster2}, Wang-Landau \cite{claudio}, PT and ST with free-energy weights \cite{fiore8,fiorejcp,fiore10,alexandra}. Therefore, they constitute relevant benchmarks for our purposes. The BL and ALG are also important examples, taking into account their more complex phase diagrams, including  liquid phases with distinct structures, regions of unusual behaviors (density and diffusion anomaly lines)  and also dynamic transitions \cite{fiore-m,fiore-m2}. In the case of ALG, an extra advantage arises, due to the existence of two phase coexisting lines, between gas and liquid phases. Hence, the ALG works as a double checking of reliability 
of   our proposals. For instance, we focus on  the regime of low temperatures, in which  strong first-order phase transitions separate coexisting phases.  Recently a general approach for discontinuous transitions has been proposed \cite{fioreprl,fiorejcp2}, in which thermodynamic quantities are described by a general function, allowing to achieve  all relevant informations
by studying rather small system sizes for some control parameters. Thus, its combination with a proper usage of ST can provide us  a powerful approach to deal with discontinuous transitions with rather low computational cost.

Henceforth, the analysis of all cases will show that the approximated weights work properly (and hence lead to correct results) in the regime of large
replica numbers $R$  for an appropriate choice of temperature sets \cite{alexandra}. On the other hand, for sufficiently small $R$ its performance is strongly reduced and the system does not visit properly the distinct coexisting regions.  Finally we extend for the approximated weights, a simple
protocol for obtaining proper temperature sets initially proposed for the free-energy weights \cite{alexandra}.

\section{Simulated Tempering and approximated  weights}

The basic idea of the ST concerns with the fact that the system temperature $T$ can assume different values 
between the extreme values $T_1$ and $T_R$, where 
$R$ is the replica number.  The MC simulation is defined as follows: 
In the first part, starting at
a given temperature  $T_{i}$ within 
the set ${\mathcal T}_R\equiv \{T_1,...,T_R\}$ 
(in all cases we started from $T_R$), 
a given site of the lattice is randomly chosen and its
variable is changed (among all possibilities)
according to the Metropolis
prescription $\min \{ 1, \exp(\beta_{i} \Delta {\cal H})\} $,
where $\Delta {\cal H}
 (\sigma)$ denotes the energy difference between the "new" and
"old" configurations and $\beta_i=1/k_{B}T_i$.  After repeating above 
dynamics a proper number of realizations (here $L^2$ random choices
are considered) in the second part
the temperature exchange ($T_{i} \rightarrow T_{j}$) occurs with the following probability 
\begin{equation}
  p_{i \rightarrow j} = \min \{ 1, \, \exp[(\beta_{i} - \beta_{j}){\cal H}
(\sigma) + (g_{j} - g_{i})] \},
  \label{pt-weight}
\end{equation}
where $g_i$ is the weight associated with the  temperature $T_i$ and ${\cal H}(\sigma)$ is the system Hamiltonian. For a uniform sampling, the weights should be proportional to the free-energy $f_i$ given by $g_{i} = \beta_{i} \, f_{i}$ \cite{pande}. Since the evaluation of  $f$ is not an 
easy task, alternative calculations of weights have been 
proposed \cite{ma,maJCF,sauerwein,fiore11}.
The simplest proposal \cite{pande} estimates the $g$'s according to the 
following approximated formula
\begin{equation}                                                               
  g_{j} - g_{i} \approx (\beta_{j} - \beta_{i}) (U_{j}+U_{i})/2,               
  \label{pande}                                                            
\end{equation}
with $U_i = \langle {\cal H}_i \rangle$ ($i=1, 2, \ldots, R$) denoting 
the average system energy at $T_i$. Thus, from Eq.(\ref{pande}) the 
weights are estimated from simple and direct standard numerical 
simulations. Here we give a further step by analyzing them by 
inspecting two crucial points: their dependence on the replica 
number $R$ and on the set of temperatures ${\mathcal T}_R$. 
In order to scrutinize them, we compare  numerical results at 
the phase coexistence points
for distinct  $R$'s, with temperature schedules estimated as 
proposed in Ref. \cite{alexandra} and described as follows: 
starting from a fixed $T_1$ we choose the next $R-1$ 
temperatures $T_2 < T_3 < \ldots < T_R$ in such a  way that 
the resulting exchange 
frequencies $f_{i+1, i}$ between any two successive temperatures 
$T_i$ and $T_{i+1}$  are all equal to some value 
specified $f_{i+1, i}=f$. We define $f_{i+1, i}$ as 
the ratio of  the number of exchanges between $T_i$ 
and $T_{i+1}$ to the total Monte Carlo steps $N_{MC}$. 
Note that from this recipe the highest temperature $T_{R}$ 
becomes automatically obtained. The efficiency of such 
achieved set ${\mathcal T}_R$ is verified by means of 
standard tests, where in the case of first-order transitions, 
the tunneling between the coexisting phases and convergence 
to the steady state starting from a non-typical initial 
configuration constitute proper efficiency measures.  More
specifically, the existence of full trapping in a given phase or even temperature
changes that do not allow the system to visit properly the coexisting phases will imply
in  thermodynamic averages   marked by no changes or abrupt variations
(see e.g Figs. 2 and 3 for $f=0.02$ and $0.021$, respectively).
Such points can be understood by recalling the ideas 
from Ref. \cite{fioreprl,fiorejcp2}, when the system 
close to the phase coexistence have typical thermodynamic 
quantities, like energy and order parameter,   well 
described by the following general expression
\begin{equation}                                                              
W(y)\approx(b_1 + \sum_{n=2}^{\mathcal{N}} b_n \, \exp[-a_n y])/(1 + \sum_{n=2}^{\mathcal{N}} c_n \, \exp[-a_n y]),                             
\label{eq1}                                                                    
\end{equation}
where for $\mathcal{N}$ coexisting phases, $y$ denotes the ``distance'' to the coexistence point $\xi^{*}$
 given by $y = \xi - \xi^{*}$. The coefficients $c_n$'s and $b_n$'s are related to derivatives of the free 
energies $f_n$ of each phase $n$  with respect to parameter  $\xi$ 
reading $\partial f_n/\partial \xi$ \cite{fiorejcp2}. In the case 
of two phase coexistence ($\mathcal{N}=2$) Eq. (\ref{eq1}) acquires the following way 
$W(y)=(b_1+b_2\exp[-a_2 y])/(1+c_2\exp[-a_2 y])$ and hence only four parameters
are necessary to determine the whole function. 
In other words, according 
to Eq. (\ref{eq1}), numerical simulations (of a given system) for known $L$ and  
control parameter sets $\xi=\xi_0$ (like chemical potential and temperature) 
will provide 
a well defined value for thermodynamic quantities $W_{0}^*=W(L,y_0)$,
where $y_0=\xi_0-\xi^*$. Note that at the phase 
coexistence point $y=0$, the quantity $W$ reads 
\begin{equation}
W_{0}=\frac{b_1+\sum_{n=2}^{\mathcal{N}} b_n}{1+ \sum_{n=2}^{\mathcal{N}} c_n }
\end{equation}
for all $L$'s. Hence, different curves of $W$ should cross at the coexistence point. 
However, as $W_{0}$ and $W_{0}^*$ are verified only
for dynamics that visit properly the distinct phases (e.g. 
one flip algorithms lead to strong hysteresis at low $T$'s and results do not obey Eq. (\ref{eq1}) ).
In the case of tempering methods, the achievement of results not following Eq. (\ref{eq1})
indicates that  ${\mathcal T}_R$ is not proper. Typically, low $T_i$'s (including
the extreme $T_R$)  provide high temperature exchanges, but
the system is not able to cross energy barriers.  Hence $f$ should be decreased, in order to raise the 
temperatures. In contrast, a very high ${\mathcal T}_R$  
(obtained from a very low $f$) may be responsible for few frequent 
 exchanges and hence poor averages are obtained. So that $f$ should be raised, in order to increase 
the temperature changes. An intermediate optimal frequency $f_{opt}$ is expected to satisfy the above 
points and hence it may lead to 
frequent  tunneling between the phases and a faster convergence 
to the steady state.  Some remarks are needed: Although our present recipe 
provides a reliable temperature set, it does not exclude the existence of other
optimized choices presenting distinct frequencies between adjacent temperatures.
Also, other efficient sets (ranged from $T_1$ and $T_R$) can be obtained. 
For simplicity, we have considered such   equal frequencies proposal.

In this work, we will test this procedure for distinct $R$'s, 
in order to verify the replica number influence in the results. 
It is worth 
mentioning that although some numerical work is necessary to 
find $f_{opt}$, in practice such process demands relative 
short simulations. Hence, the search for 
the corresponding optimal ${\mathcal T}_R$ is not 
a computationally time consuming procedure.

\section{Models}

\subsection{BEG and BC models }

The BEG  model Hamiltonian \cite{BEGMODEL} is given by 
\begin{equation}                                                               
  {\cal H} = -\sum_{<i,j>} (J \sigma_{i} \, \sigma_{j} + K \sigma_{i}^{2} \sigma_{j}^{2}) +D \sum_{i}  \sigma_i^2,  
  \label{e3}                                                                     
\end{equation}
where $\sigma_i=0$, if  the site $i$ has null spin  and $\pm 1$ if $i$ has up and down values, respectively. 
The parameters  $J$ and $K$ are interaction energies and $D$  denotes the crystalline field. The BC model corresponds to the ${\bar K}\equiv K/J=0$ case. In order to compare with previous results \cite{cluster2,fiorejcp,alexandra} 
we also consider the  ${\bar K}=3$ case, in  which the system displays ferromagnetic  and paramagnetic phases. All phases  can be characterized in terms of two order-parameters $q=\langle \sigma_{i}^{2}\rangle $  and $m=\langle \sigma_{i}\rangle$. At low temperatures, all transitions are first-order and yield close  to the $T=0$ the value ${\bar D}^{*}=z({\bar K}+1)/2$, where $z$ is the coordination number of the lattice. In both model studies, we consider a 
two-dimensional square lattice in which $z$ reads $4$.

\subsection{Bell-Lavis model}

The  Bell-Lavis (BL) model is defined on a triangular lattice ($z=6$) where each site is described by two kinds of variables, namely occupational ($\sigma$) and orientational ($\tau$) states. The former takes the values $\sigma_i=0$ or $1$, whenever the site $i$ is empty  or occupied by a water molecule, respectively. The latter describes the possibility of forming hydrogen bonds and reads $\tau_{i}^{ij} = 0$ or $\tau_{i}^{ij} = 1$ when the arm is inert or bonding, respectively. 
Two  molecules interact via  van der Waals and hydrogen bond energies provided they are adjacent and  point out their arms to each other ($\tau_{i}^{ij}\tau_{j}^{ji} = 1$), respectively. The BL model is given by the following Hamiltonian  
\begin{equation}                                                               
  {\mathcal H} = -\sum_{<i,j>} \sigma_{i} \, \sigma_{j} \, (\epsilon_{hb} \, \tau_{i}^{ij} \, \tau_{j}^{ji} + \epsilon_{vdw}) - \mu \sum_{i} \sigma_{i},                                     
  \label{hambl}                                                                  
\end{equation}
where $\epsilon_{vdw}$ and $\epsilon_{hb}$ are the van der Waals and hydrogen bonds interaction energies, respectively,  and $\mu$ is the chemical potential. We also compared results with those obtained from PT and ST with  free-energy weights \cite{fiore11,alexandra}. For instance,  we consider $\zeta=0.1$ (where  $\zeta=\epsilon_{vdw}/\epsilon_{hb}$), in which  the system presents a gas and two distinct liquid phases, named low-density-liquid (LDL) and  high-density-liquid (HDL)  \cite{bell,fiore-m}. In the gas and HDL phases, the lattice is empty and it is almost filled by molecules  respectively. On the other hand, the LDL phase presents an intermediate  density close to $\rho=2/3$ and exhibits an ordering structure  (honeycomb like geometry), signed by a maximum  density of hydrogen bonds per molecule.  At ${\bar T}=0$
(where ${\bar T \equiv T/\epsilon_{hb}}$), both transitions are first-order and occurs at ${\bar \mu}^{*} = -3 \, (1+\zeta)/2$ and  ${\bar \mu}^{*} = -6 \, \zeta$, respectively. For ${\bar T} \neq 0$ and distinct $\zeta$'s the former phase transition  remains first-order, but the latter  becomes critical \cite{fiore-m}. For $\zeta=0.1$, 
the second-order and first-order lines meet in a tricritical point.

\subsection{Associating lattice-gas (ALG) model}

Similarly to the BL, the ALG model is also described by an occupation ($\sigma$) and orientation ($\tau$) states. But an important difference from the BL is that an energetic punishment exists when a hydrogen bond is not formed. More specifically, two first neighbor molecules have an interaction energy of $-v$ ($-v + 2 u$) if there is (there is not) a hydrogen bond between them. The Hamiltonian system reads
\begin{equation}                                                               
{\cal H} = 2 u \sum_{<i,j>} \sigma_{i} \sigma_{j} [(1 - v/(2 u)) -             
\tau_{i}^{ij} \tau_{j}^{ji}] - \mu \sum_{i}\sigma_i.                           
\end{equation}
The ALG also presents a gas, LDL and HDL phases, with densities $\rho=3/4$ 
and $1$, respectively \cite{algs}. Another relevant distinction from the
 BL model is that here both gas-LDL and LDL-HDL transitions remain first-order 
for ${\bar T}\equiv T/v \neq 0$
and end at respective critical points joined by a critical line 
separating a structured liquid from a disordered fluid phase. At ${\bar T}=0$, 
the discontinuous transitions occur at ${\bar \mu} \equiv \mu/v=-2$ (gas-LDL) 
and ${\bar \mu}=-6+8u/v$ (LDL-HDL). As in the BL model, $\rho$ is 
a proper parameter for the gas-LDL transition, but not 
for other phase transitions. Alternative 
$\phi$'s are $\phi = (4\rho -3)$ or also $\phi=(2\rho_{hb}-3)$, being $\rho_{hb}$ the hydrogen bonds density, taking
the values  $\rho_{hb}=3/2$ and $1$ for LDL and HDL phases, respectively. 
Here, we  focus on the temperature ${\bar T}=0.30$, in which results 
from the PT show that the former and latter transitions take place at 
${\bar \mu^{*}} = -1.9986(2)$ and 
${\bar \mu^{*}}=2.0000(2)$, respectively \cite{fioreprl}.

\section{Numerical results}

\begin{figure}
\setlength{\unitlength}{1.0cm}
\includegraphics[scale=0.35]{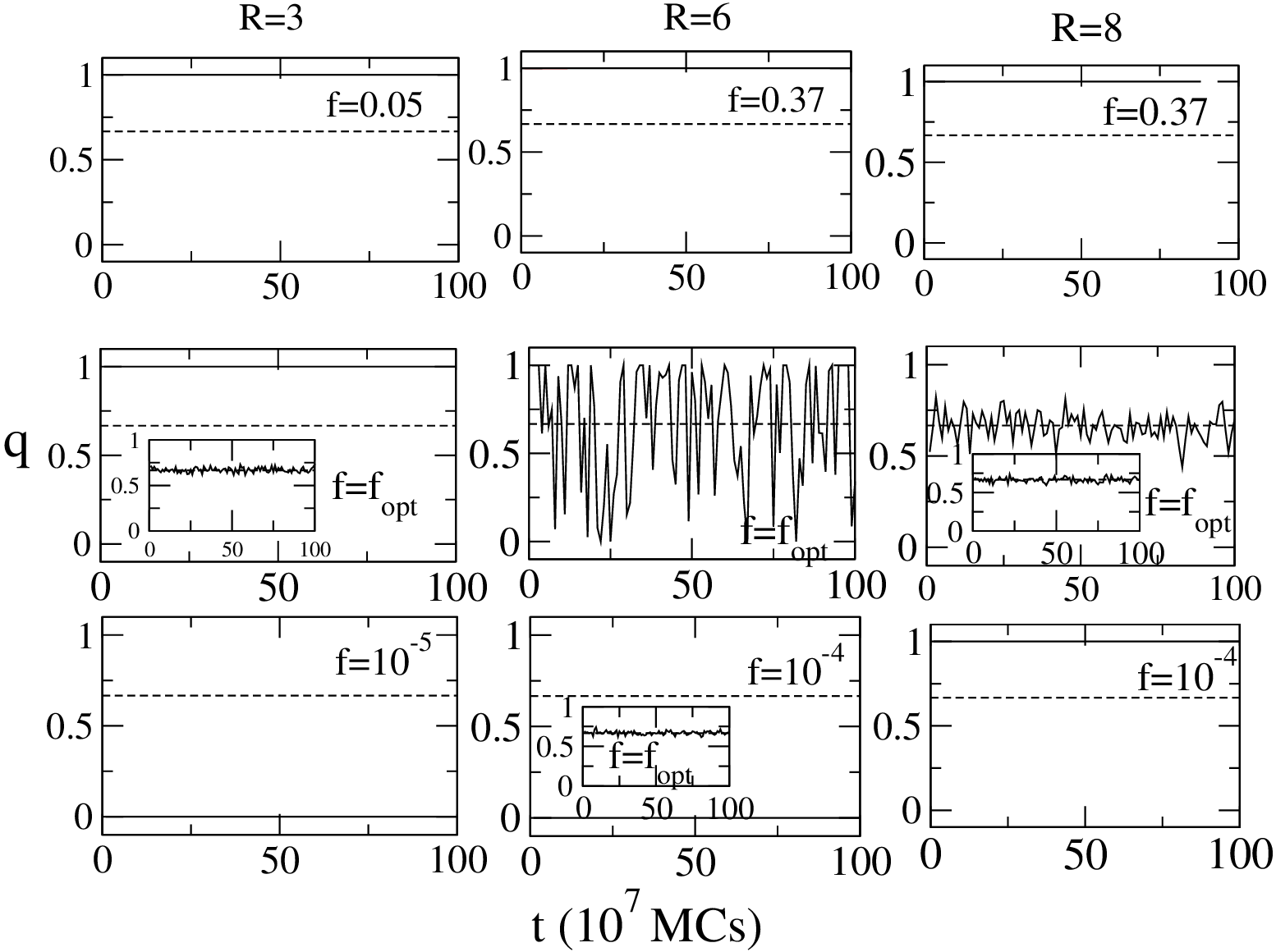}                
\caption{Order parameter $q$ as a function of $t$ for the BEG 
model at $T=0.5$ for $L=20$. Distinct values of $f$'s and $R$'s were
considered at the phase coexistence $D^{*}=8.0000(1)$, with temperature sets obtained from Ref. \cite{alexandra} but using the approximated weights. The dashed lines correspond to steady value  $q_0=2/3$. The insets correspond to the results from free-energy weights.}
\label{fig1}
\end{figure}

\begin{figure}
\setlength{\unitlength}{1.0cm}
\includegraphics[scale=0.35]{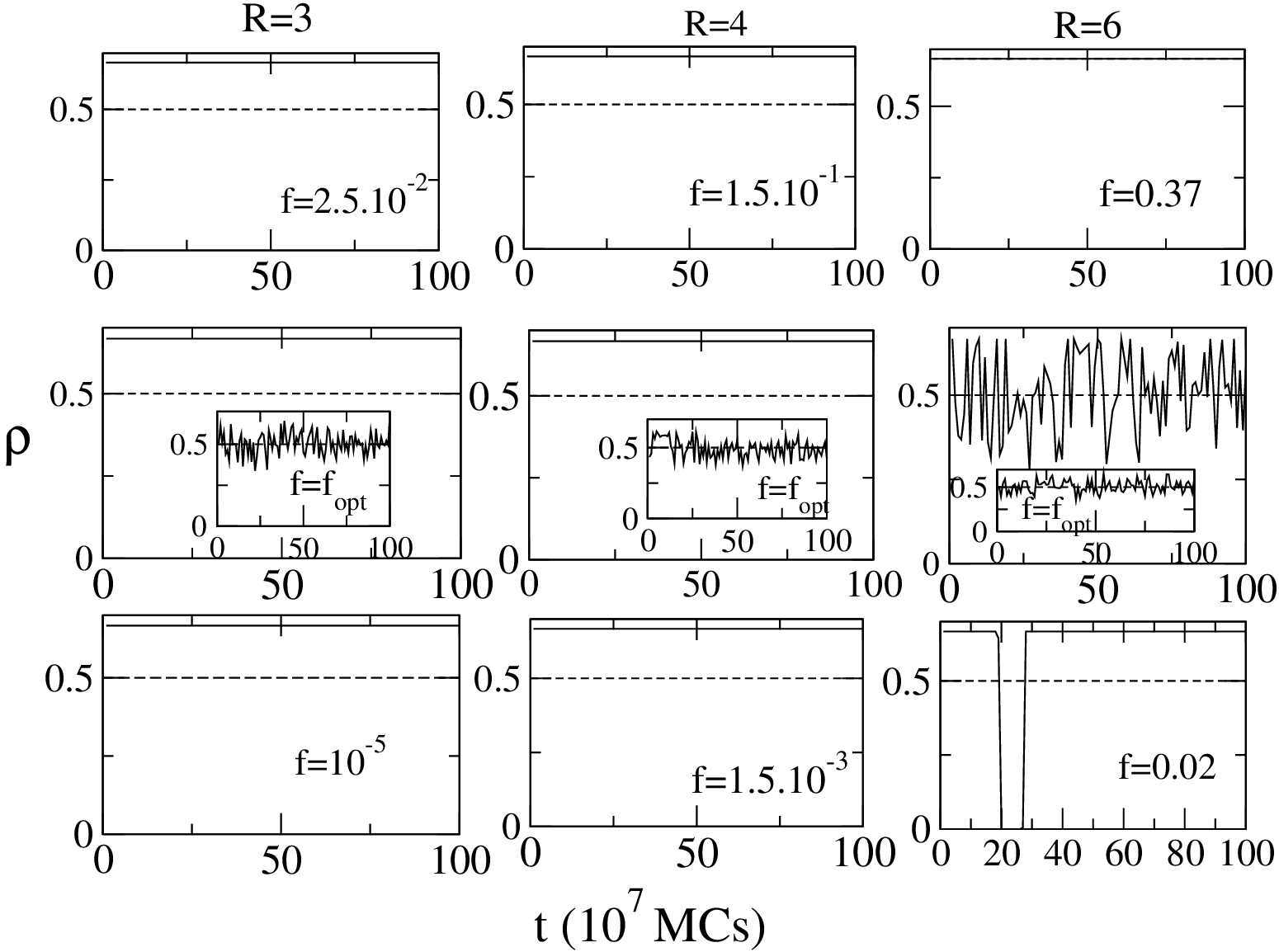}                
\caption{For the BL model and $T=0.1$, $\rho$ vs $t$ for the BL model at the phase coexistence $\mu^{*}=-1.6500(1)$ 
for distinct $R$'s and $f$'s, with temperature sets obtained from Ref. \cite{alexandra} but using  the approximated weights. 
Numerical simulations have been carried out for $L=18$.
The dashed lines denote the steady density $\rho_0=1/2$ and the insets 
correspond to the free-energy weights. }
\label{fig2}
\end{figure}

 Numerical simulations were carried out for several system sizes $L$ 
and control parameters. In all cases, numerical simulations start from
a fully ordered initial configuration and $3.10^6$ MC steps have been used
to ``equilibrate'' the system. After the transient,  we have 
used $3.10^6$ MC steps for evaluating the weights and steady analysis
start from the largest temperature $T_R$. Except Fig. \ref{fig4}$(a)$, 
all quantities have been evaluated after the transient regime.
 From now on, we will replace all reduced quantities
(${\bar T}, {\bar H},{\bar D}, {\bar \mu}$) for  ($T, H, D, \mu$).
In order to achieve a global idea about approximated weights and how they compare with free-energy ones, in the first analysis we show the tunneling between coexisting phases for the BEG and BL models, respectively by taking the temperature schedules ${\mathcal T}_R$ (for distinct $R$'s) obtained from free-energy weights \cite{alexandra}. For low temperature values, $T_1=0.5$ (BEG) and 
$0.1$ (BL), the coexistence points yield at ${\bar D^{*}}=8.0000(1)$ and ${\bar \mu^{*}}=-1.6500(1)$, respectively. In these points, according to Eq. (\ref{eq1}) a correct system sampling should give steady  equilibrium values consistent with  $q_{0}  \approx 2/3$ and $\rho_{0}  \approx 1/2$ for the BEG and BL, respectively. The former can be understood by recalling that two ferromagnetic phases ($q \approx 1$) coexist with one paramagnetic phase ($q \approx 0$). Since at the phase coexistence their weights are equal (1/3), we have $q_{0} \approx 2/3$ for all $L$'s. A similar reasoning holds for the latter model, in which  the LDL phase has density $\rho \approx 2/3$ (with degeneracy $3$) and 
coexist with the gas phase, in such a way that at ${\bar \mu^*}$ 
the value $\rho_0 \approx 1/2$ is verified. For this study, three values of $R$ and frequencies were considered, whose temperature sets and distinct $f$'s are obtained from free-energy weights and are shown in Tables \ref{table1}, \ref{table2} and \ref{table3} (BEG) and Tables \ref{table4},\ref{table5} and \ref{table6} (BL).
Results are summarized in Figs. \ref{fig1} and \ref{fig2} for $L=20$ and $L=18$,
respectively.  As discussed previously, 
in all cases extremely low or large frequencies provide no precise results, signaling trapping 
in a given phase ($q=0$ or $1$ for the BEG and $\rho=0$ and $2/3$ for the BL), 
due to low $T$'s or hardly exchanges.  However, a similar result is verified for the 
intermediate $f$ and the lowest  $R=3$,
in such a way that only for intermediate frequencies  ($f=f_{opt}$) when  $R=6$ and $R=8$, the crossing between coexisting phases is verified. However, for $R=6$  the averages deviate greatly from its mean value $2/3$, indicating that despite  tunneling between coexisting phase occurs, it is not efficient. Improved results are  obtained  only for $f_{opt}$ and $R=8$, in which the tunneling is more frequent. These results present stark difference with  those obtained by taking the free-energy weights (insets), in which for all replica numbers (including the lowest $R=3$ case), it is possible to find an optimal $f_{opt}$, ensuring proper tunneling between phases. Thus, the first evidence suggests that  the choice of the replica number $R$ is more important with approximated weights than with the
free-energy ones.
\begin{table}
\begin{ruledtabular}
\begin{tabular}{cccc}
f & $5 \times 10^{-2} $ &  $6 \times 10^{-4}$ ($f_{opt}$) & $10^{-5}$  \\
\hline
$T_1$ & 0.50 & 0.50 & 0.50    \\
$T_2$ & 1.35 & 1.60 & 1.82    \\
$T_3$ & 1.70 & 2.05 & 2.33    \\
\end{tabular}
\end{ruledtabular}
\caption{Temperature sets ${\mathcal T}_{R=3}$, for BEG model, considering 
the frequencies $f$ obtained 
from free-energy weights \cite{alexandra}.\label{table1}}
\end{table}

\begin{table}
\begin{ruledtabular}
\begin{tabular}{cccc}
f & $ 0.37$ &  $0.11$ ($f_{opt}$) & $10^{-4}$  \\
\hline
$T_1$ & 0.50 & 0.50 & 0.50    \\
$T_2$ & 1.12 & 1.25 & 1.70    \\
$T_3$ & 1.24 & 1.55 & 2.14    \\
$T_4$ & 1.34 & 1.78 & 3.20    \\
$T_5$ & 1.42 & 1.95 & 4.34    \\
$T_6$ & 1.50 & 2.06 & 6.10    \\
\end{tabular}
\end{ruledtabular}
\caption{Temperature sets ${\mathcal T}_{R=6}$, for the
BEG model, considering distinct frequencies $f$ obtained from 
free-energy weights \cite{alexandra}.\label{table2}}
\end{table}

\begin{table}
\begin{ruledtabular}
\begin{tabular}{cccc}
f & $ 0.37$ &  $0.21$ ($f_{opt}$) & $10^{-4}$  \\
\hline
$T_1$ & 0.50 & 0.50 & 0.50    \\
$T_2$ & 1.12 & 1.16 & 1.70    \\
$T_3$ & 1.24 & 1.39 & 2.14    \\
$T_4$ & 1.34 & 1.56 & 3.20    \\
$T_5$ & 1.42 & 1.72 & 4.34    \\
$T_6$ & 1.50 & 1.85 & 6.10    \\
$T_7$ & 1.58 & 1.96 & 6.60    \\
$T_8$ & 1.64 & 2.03 & 7.00    \\
\end{tabular}
\end{ruledtabular}
\caption{Temperature sets ${\mathcal T}_{R=8}$, for the BEG model, 
considering distinct frequencies $f$ obtained 
from free-energy weights \cite{alexandra}.\label{table3}}
\end{table}

\begin{table}
\begin{ruledtabular}
\begin{tabular}{cccc}
f & $ 0.025$ &  $2 \times 10^{-4}$ ($f_{opt}$) & $10^{-5}$  \\
\hline
$T_1$ & 0.10 & 0.10 & 0.10    \\
$T_2$ & 0.25 & 0.32 & 0.35    \\
$T_3$ & 0.33 & 0.43 & 0.48    \\
\end{tabular}
\end{ruledtabular}
\caption{Temperature sets ${\mathcal T}_{R=3}$, 
for the BL model, considering distinct frequencies $f$ obtained from 
free-energy weights \cite{alexandra}.\label{table4}}
\end{table}

\begin{table}
\begin{ruledtabular}
\begin{tabular}{cccc}
f & $ 0.15$ &  $0.01$ ($f_{opt}$) & $1.5 \times 10^{-3}$  \\
\hline
$T_1$ & 0.10 & 0.10 & 0.10    \\
$T_2$ & 0.20 & 0.27 & 0.29    \\
$T_3$ & 0.25 & 0.34 & 0.39    \\
$T_4$ & 0.29 & 0.43 & 0.50    \\
\end{tabular}
\end{ruledtabular}
\caption{Temperature sets ${\mathcal T}_{R=4}$, for the BL model, 
considering distinct frequencies $f$ obtained from free-energy weights \cite{alexandra}.\label{table5}}
\end{table}

\begin{table}
\begin{ruledtabular}
\begin{tabular}{cccc}
f & $ 0.37$ &  $0.07$ ($f_{opt}$) & $0.02$  \\
\hline
$T_1$ & 0.100 & 0.10 & 0.10    \\
$T_2$ & 0.200 & 0.23 & 0.25    \\
$T_3$ & 0.230 & 0.30 & 0.32    \\
$T_4$ & 0.260 & 0.32 & 0.41    \\
$T_5$ & 0.285 & 0.39 & 0.47    \\
$T_6$ & 0.312 & 0.43 & 0.60    \\
\end{tabular}
\end{ruledtabular}
\caption{Temperature sets ${\mathcal T}_{R=6}$, for the BL model, 
considering distinct frequencies $f$ obtained 
from free-energy weights \cite{alexandra}.\label{table6}}
\end{table}

 
\begin{figure}
\setlength{\unitlength}{1.0cm}
\includegraphics[scale=0.35]{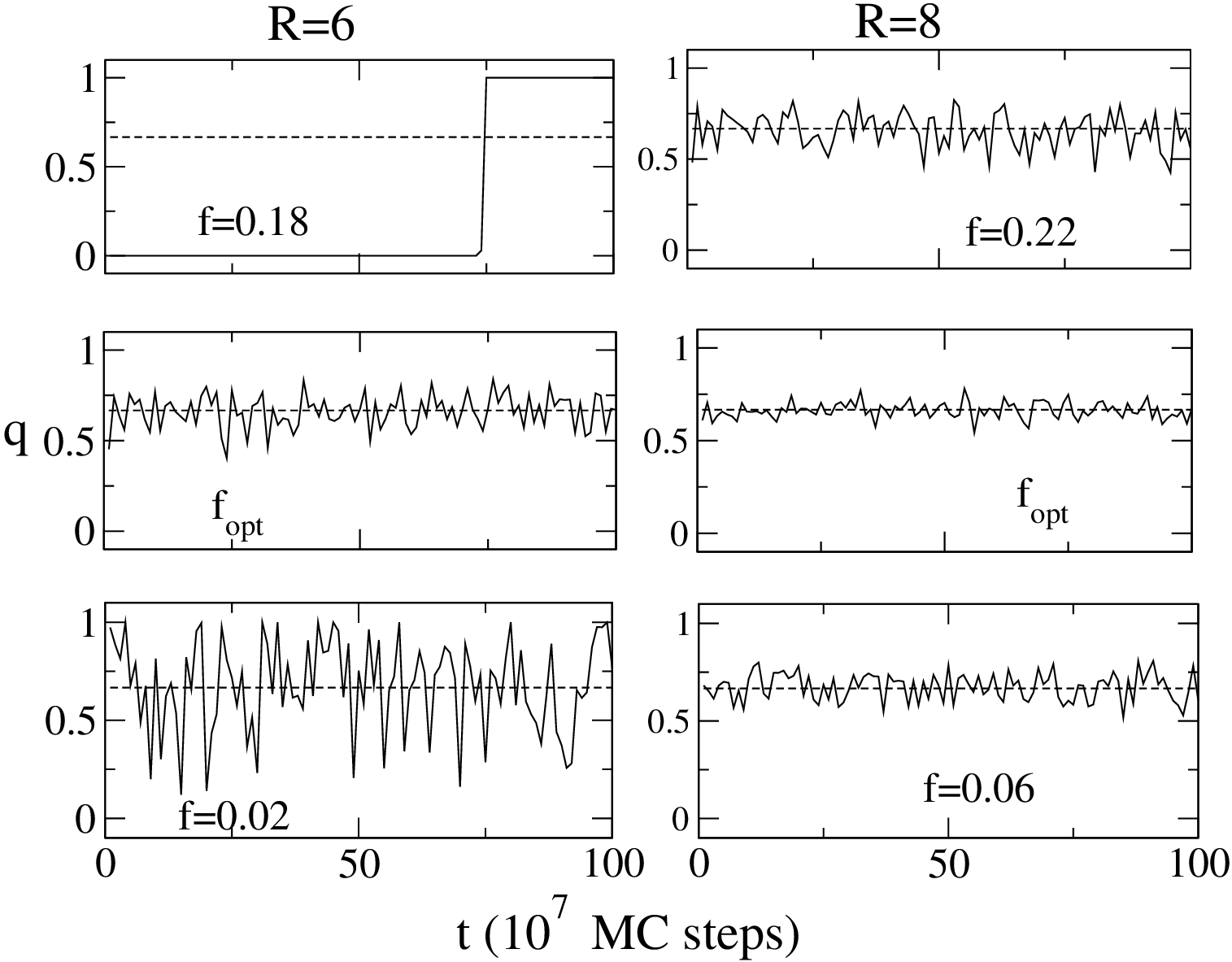}                
\caption{For the BEG model, $T=0.5$ and $L=20$, $q$ versus $t$ 
for distinct  $f$'s and  $R$'s at the phase coexistence 
$D^{*}=8.0000(1)$, whose sets are obtained from approximated weights. The dashed 
lines correspond to steady value  $q_0=2/3$.}
\label{fig3}
\end{figure}

\begin{figure}
\setlength{\unitlength}{1.0cm}
\includegraphics[scale=0.34]{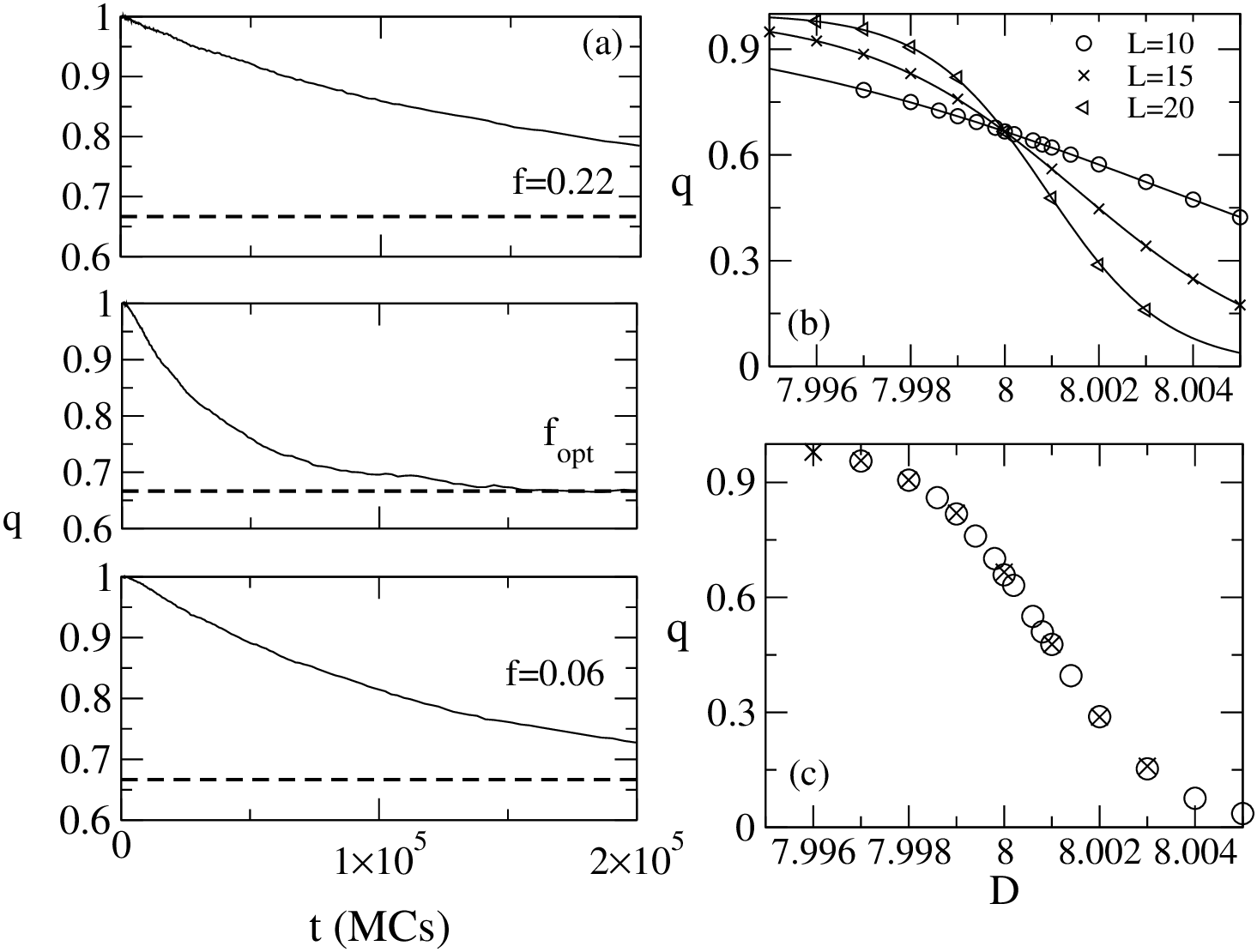}                
\caption{For the BEG model, panel $(a)$ shows the time decay of the order-parameter $q$ for 
distinct $f$'s at the phase coexistence $D^{*}=8.0000(1)$ 
for $R=8$ and $L=20$. In $(b)$  $q$ vs $D$ for distinct $L$'s, with numerical results 
obtained from ${\mathcal T}_{R=8}$ for $f_{opt}$.
The continuous lines denote results  for $q$ obtained from Eq. (\ref{eq1}). In $(c)$ we 
compare results with those obtained from approximated (stars) and free-energy (circles) weights. }
\label{fig4}
\end{figure}

The results  for the BL shown in Fig. \ref{fig2} reinforce this idea. Only for the largest case $R=6$ with $f=f_{opt}$, some tunneling between the gas and LDL phases occurs. As for the BEG model, free-energy weights provide reliable results for 
all $R$'s (insets). The crucial importance  of $R$ is understood as follows: For  two arbitrarily temperatures $T_i$ and $T_j$ the exchange frequency obtained from the approximated weights is lower than those from the free-energy ones, whose difference becomes more pronounced for the optimal choices. For example, for the BEG and $R=3$ the exchange between $T_1=0.50$ and $T_2=1.60$ (Table I)
is performed with frequency $f_{opt}=6.10^{-4}$ when one takes the free-energy
weights. For the approximated weights, it reads $10^{-6}$.
An efficient performance of the ST 
(e.g. the system visiting properly distinct coexisting phases) depends 
not only on frequent exchanges but also a reliable estimation of all temperature schedules, 
including the extreme temperature $T_R$. Thus, the compromise between these points is reached  
(by taking the approximated weights) only for larger $R$'s.
For example, 
in such regime (exemplified here for $R=8$) the exchange 
between $T_1=0.50$ and $T_2=1.16$ (Table III) is  performed 
with a considerable larger frequency, reading  $f=0.21$ and $f=0.015$ 
when one takes the free-energy and approximated weights, respectively.
We remark that frequency $0.015$ is larger than
the value $10^{-6}$ obtained between $0.5$ and $1.60$ (approximated
weights).  
For the BEG model, in Fig. \ref{fig3} we confirm these points by obtaining temperature schedules ${\mathcal T}_R$'s  (for $R=6$ and $R=8$) with frequencies evaluated from the approximated weights, instead of those available from free-energy values \cite{alexandra}. As in previous cases,  three frequencies have been considered, with temperature sets  shown in Tables \ref{table7} and \ref{table8} for $R=6$ and $R=8$, respectively. As in  Fig. \ref{fig1},   larger  $f$'s provide  more frequent exchanges, but it is  not efficient since the obtained $T_R$ is low. On the contrary, lower $f$ gives a larger $T_R$ but temperatures exchanges are hardly performed (clearly shown for the lowest $f=0.021$ when $R=6$). By increasing  $R$, it becomes  possible to obtain an intermediate frequency $f_{opt}$ that fulfills the two requirements above for an efficient performance (larger $T_R$ and frequent temperature exchanges). 
\begin{table}
\begin{ruledtabular}
\begin{tabular}{cccc}
f & $ 0.18$&  $0.06$ ($f_{opt}$) & $0.02$  \\
\hline
$T_1$ & 0.50 & 0.50 & 0.50    \\
$T_2$ & 1.02 & 1.10 & 1.15    \\
$T_3$ & 1.31 & 1.48 & 1.58    \\
$T_4$ & 1.53 & 1.77 & 1.88    \\
$T_5$ & 1.68 & 1.97 & 2.08    \\
$T_6$ & 1.83 & 2.12 & 2.47    \\
\end{tabular}
\end{ruledtabular}
\caption{Temperature sets ${\mathcal T}_{R=6}$, for the BEG model,
considering distinct frequencies $f$ obtained from the 
approximated weights.\label{table7}}
\end{table}

\begin{table}
\begin{ruledtabular}
\begin{tabular}{cccc}
f & $ 0.22$ &  $0.10$ ($f_{opt}$) & $0.06$  \\
\hline
$T_1$ & 0.50 & 0.50 & 0.50    \\
$T_2$ & 1.00 & 1.07 & 1.10    \\
$T_3$ & 1.27 & 1.42 & 1.48    \\
$T_4$ & 1.47 &1.68  & 1.77    \\
$T_5$ & 1.63 &1.87  & 1.97    \\
$T_6$ & 1.77 &2.01  & 2.12    \\
$T_7$ & 1.89 &2.15  & 2.44    \\
$T_8$ & 1.98 &2.43  & 3.03    \\
\end{tabular}
\end{ruledtabular}
\caption{Temperature sets ${\mathcal T}_{R=8}$, for the BEG model,
considering distinct frequencies $f$ obtained from the 
approximated weights. \label{table8}}
\end{table}

 Fig. \ref{fig4}$(a)$ reinforces above ideas  by showing for the BEG model the time decay 
of the order-parameter $q$ starting from a fully occupied initial configuration for distinct $f$'s and $R=8$. Note that the optimal choice  $f_{opt}$ also ensures the faster convergence toward the steady value $2/3$. In  panel $(b)$ we extend the obtained
${\mathcal T}_{R=8}$ (for the coexistence point $D^*$) for other $D$'s and distinct system sizes.
As for free-energy case  \cite{alexandra,fiorejcp}, an unique ${\mathcal T}_R$'s  can be 
used for other control parameters and system sizes (panel $(c)$), providing to 
characterize very precisely the transition point, 
as predicted by Eq. (\ref{eq1}) by taking only small system sizes.

\begin{figure}
\setlength{\unitlength}{1.0cm}
\includegraphics[scale=0.34]{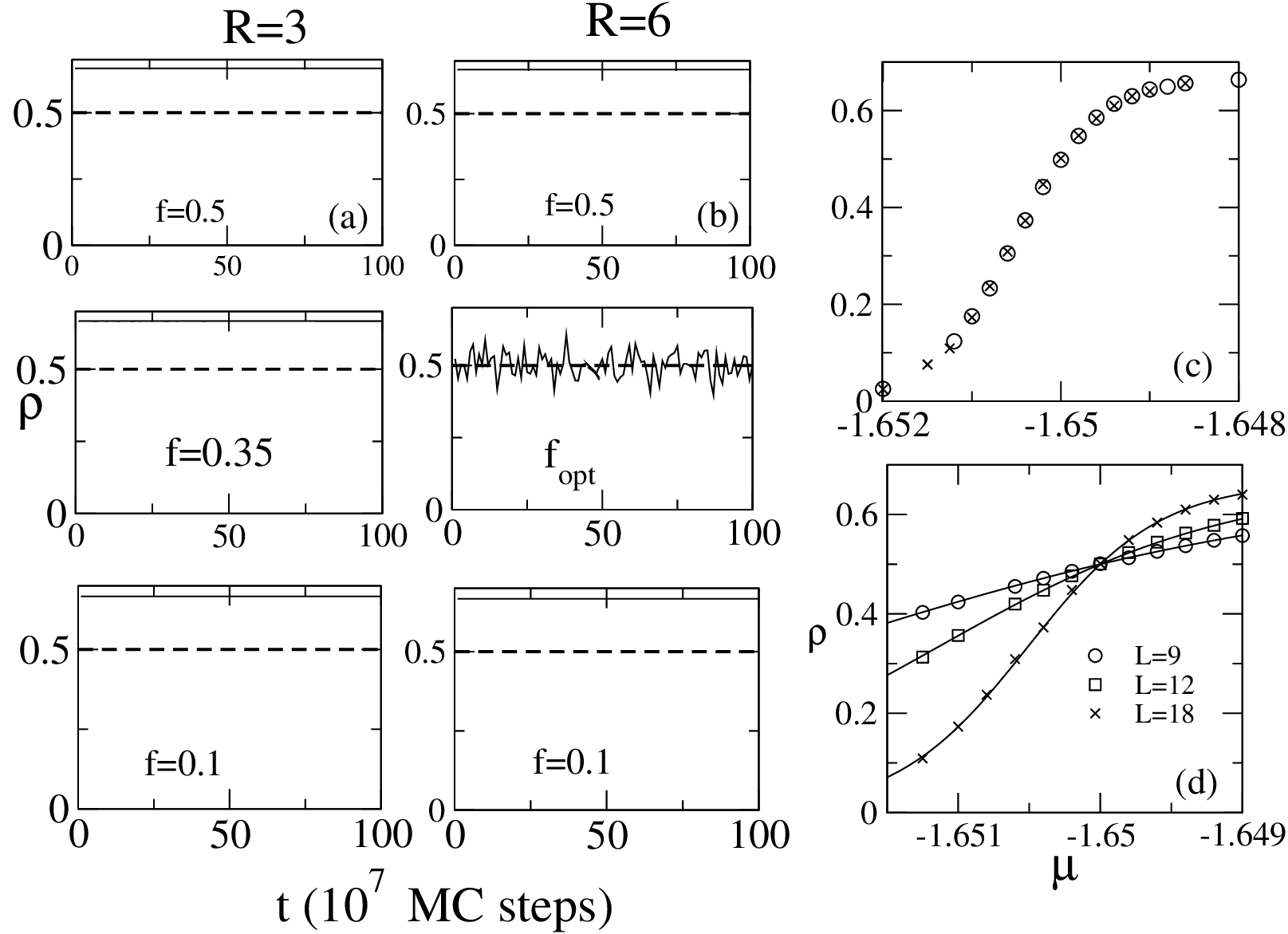}                
\caption{In $(a)$ and $(b)$, $\rho$ versus $t$ for distinct values of $f$'s and  $R$'s at the 
phase coexistence point ($T=0.1$, $\mu^{*}=-1.6500(1)$) for the BL model  and $L=18$. The dashed lines 
denote the steady density $\rho_0=1/2$. From top to bottom, intermediate temperature schedules 
read $\{0.183,0.230\}$, $\{0.200,0.266\}$ and $\{0.30,0.40\}$ ($R=3$) 
and $\{0.183,0.230,0.273,0.312,0.348\}$, $\{0.200,0.266,0.322,0.372,0.412\}$ 
and $\{0.300,0.400,0.445,0.526,0.670\}$ 
($R=6$).  In $(c)$ we compare with results obtained from free-energy weights (circles). 
In $(d)$ $\rho$ versus $\mu$ for distinct system sizes for $T=0.1$. The continuous lines denote results  
for $\rho$ obtained from Eq. (\ref{eq1}).}
\label{fig5}
\end{figure}

In Fig. \ref{fig5} we compare distinct ${\mathcal T}_R$'s for the BL model, 
by taking approximated weights. Confirming
previous results, the optimal frequency $f_{opt}$ is obtained only by increasing $R$ 
(panels $(a)$ and $(b)$) with results equivalent with the free-energy weights
(panel $(c)$).  As for the BEG model, the same ${\mathcal T}_R$ is extended to values 
of $\mu$ and system sizes, as shown in panel $(d)$ with results obeying Eq. (\ref{eq1}).

\begin{figure}
\setlength{\unitlength}{1.0cm}
\includegraphics[scale=0.34]{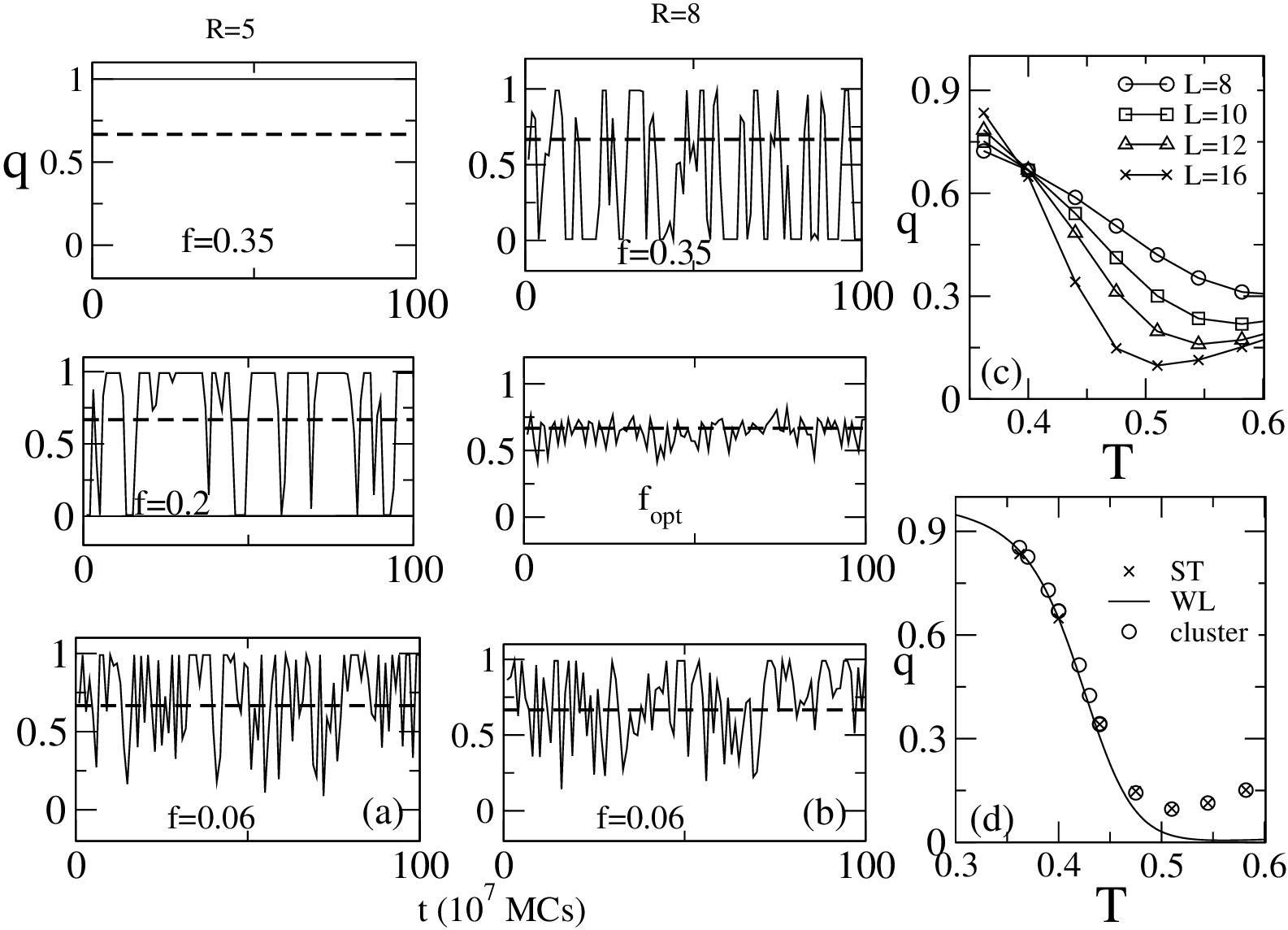}                
\caption{For the BC model, panels  $(a)$ and $(b)$, $q$ versus $t$ for distinct values of $f$'s and  $R$'s at the
 phase coexistence point ($T=0.40$, $D^{*}=1.9968$) for $L=16$. The dashed lines denote the steady 
value $q_0=2/3$.  In $(c)$ we plot $q$ vs $T$ for distinct $L$'s. The curves cross at 
the transition  point $T=0.40(1)$.  In $(d)$, we compare results obtained from the WL method 
and cluster algorithms.}
\label{fig6}
\end{figure}

Next, we analyze the BC model. 
Besides testing the previous ideas, such analysis also aims to consider  
a phase transition ruled by the temperature. Thus, it also shows the 
reliability of the whole temperature
set obtained from $f_{opt}$. We take the value $D^{*}=1.9968$, in which results from the cluster 
algorithms \cite{bouabci}, WL method \cite{claudio} and PT \cite{fiore8} predicts a phase coexistence close to 
$T=0.4$.  Results for the tunneling between coexisting phases are
considered for $L=16$,  as shown in Fig. \ref{fig6}. From top to bottom panels, temperature schedules read 
$\{0.375,0.400,0.420,0.440,0.458\}$, $\{0.350,0.400,0.440, 0.485,0.527\}$ and $\{0.310,0.400,0.481,0.560,0.650\}$ ($R=5$) and $\{0.375,0.400,0.420,0.440,0.458,0.480,0.509,0.530\}$, $\{0.350,0.400,0.440,0.485,0.527,0.570,0.619,0.680\}$ and 
$\{0.310,0.400,0.481,0.560,0.650,0.783,0.985,1.290\}$ ($R=8$).
As in previous examples,  small  $R$ leads to trapping of the system 
with the method performance improving  as $R$ increases. As it 
can be seen, the achievement of an optimal  ${\mathcal T}_R$ ensures
 tunneling and faster  convergence toward the equilibrium value. Also, the
 same  ${\mathcal T}_R$ is extended for distinct $L$'s and
 from Eq. (\ref{eq1}) we see that all curves cross at $T=0.40(1)$ (panel $(c))$. For 
 $T>0.5$, we see an unusual behavior, signaling
the large temperature regimes and the non-validity of Eq. (\ref{eq1}). 
Comparison with  cluster algorithms \cite{bouabci}
and WL results \cite{claudio} (panel $(d)$) confirms once again the main ideas that 
approximate weights can be used properly 
in order to give results equivalent with such tools.

\begin{figure}                                                                
\setlength{\unitlength}{1.0cm}                                                
\includegraphics[scale=0.34]{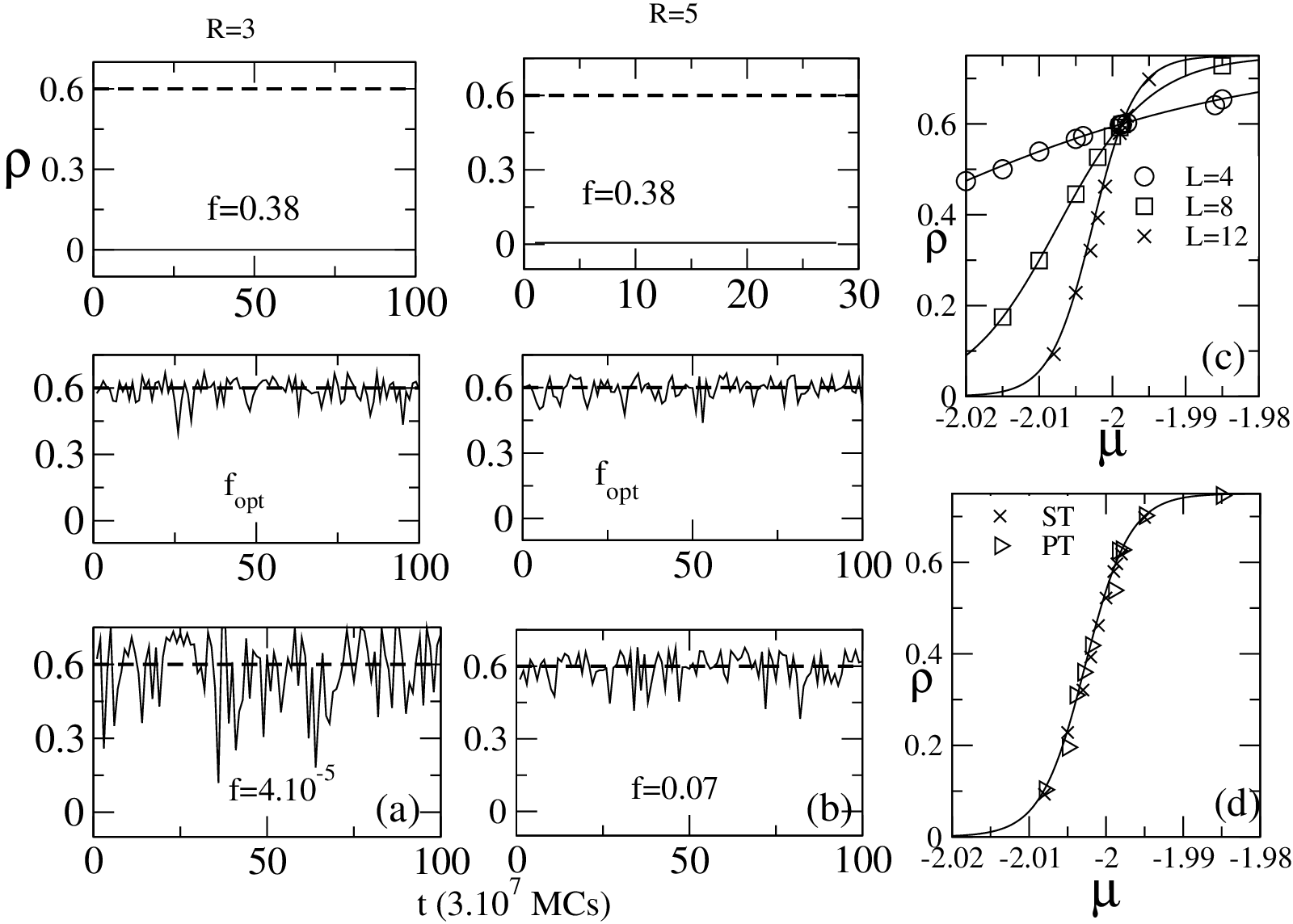}                                 
\caption{For the ALG model and $L=12$, panels $(a)$ and $(b)$ show $\rho$ versus $t$
for distinct values of $f$'s and  $R$'s at the gas-LDL phase coexistence 
point $\mu^{*}=-1.9986(1)$  and $T=0.30$. 
The dashed lines denote the steady density $\rho_0=3/5$. 
From top to bottom, intermediate 
temperature schedules read 
$\{0.32,0.34\}$, $\{0.406,0.520\}$ and $\{0.44,0.56\}$ ($R=3$) 
and $\{0.32,0.34,0.355,0.372\}$, $\{0.355,0.395,0.434,0.473\}$ 
and $\{0.38,0.45,0.53,0.653\}$ ($R=5$).  In $(c)$ $\rho$ versus $\mu$ for 
distinct system sizes. The continuous lines denote results  
for $\rho$ obtained from Eq. (\ref{eq1}). In $(d)$ we compare with results obtained
from the PT (triangles) for $L=12$.}                                                                    
\label{fig7}                                                                  
\end{figure}

\begin{figure}
\setlength{\unitlength}{1.0cm}
\includegraphics[scale=0.34]{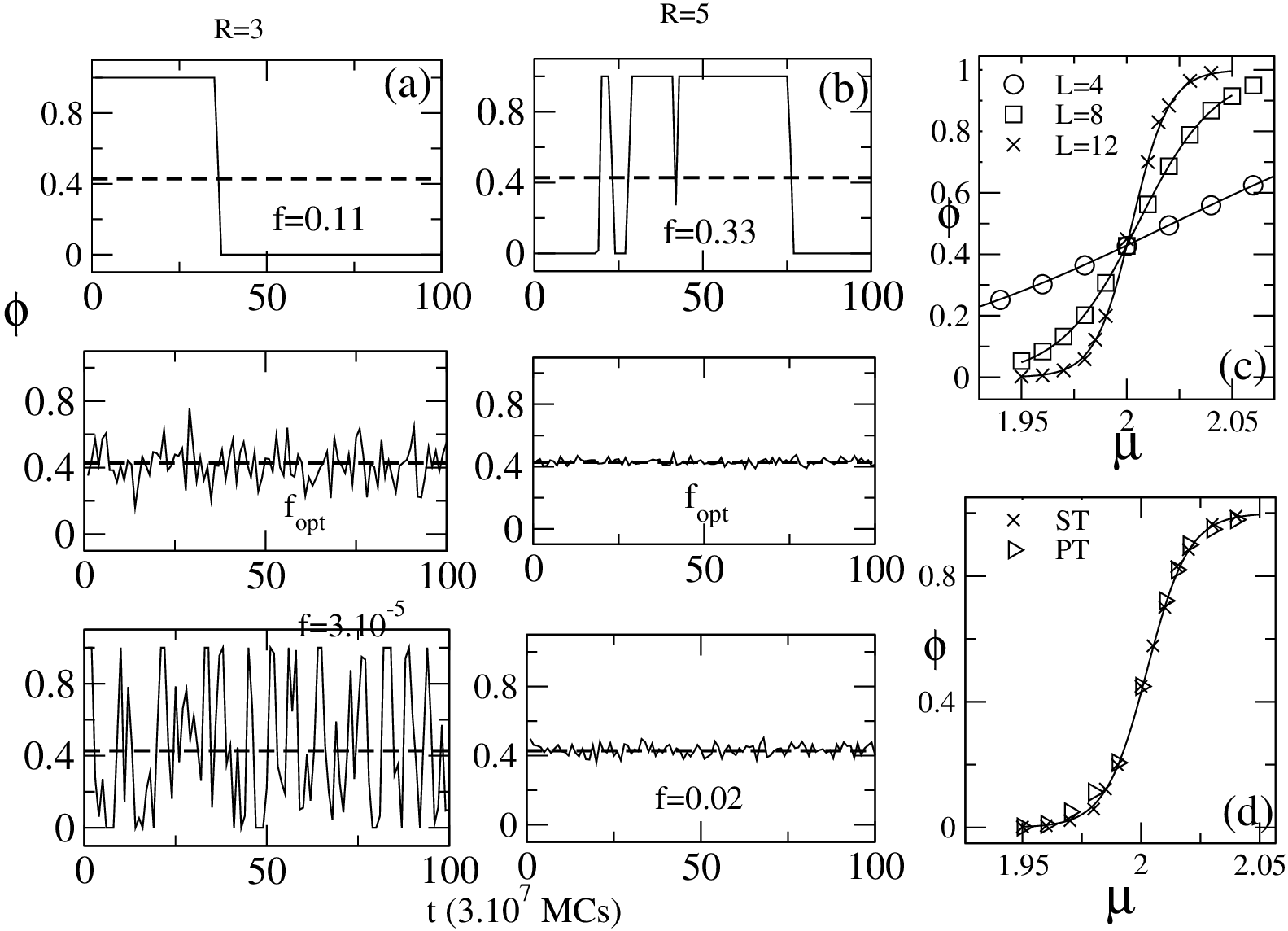}
\caption {For the ALG model and $L=12$, panels $(a)$ and $(b)$ show $\phi$ versus $t$
for distinct values of $f$'s and  $R$'s at the LDL-HDL phase coexistence 
point $\mu^{*}=2.000(1)$ and $T_1=0.30$. 
The dashed lines denote the steady
density $\phi_0=0.425$. From top to bottom, intermediate 
temperature schedules read 
$\{0.55,0.70\}$, $\{0.60,0.78\}$ and $\{0.65,0.84\}$ ($R=3$) 
and $\{0.50,0.579,0.650,0.705\}$, 
$\{0.55,0.70,0.795,0.840\}$ and $\{0.60,0.78,0.89,1.20\}$ 
($R=5$).  In $(c)$ $\phi$ versus $\mu$ for 
distinct system sizes. The continuous lines denote results  
for $\phi$ obtained from Eq. (\ref{eq1}). In $(d)$ we compare with results obtained
from the PT (triangles) for $L=12$.}
\label{fig8}
\end{figure}

For the last example, Figs. \ref{fig7} and \ref{fig8} summarize results for the ALG model
by taking  the gas-LDL and LDL-HDL coexistence points for $T=0.30$. 
In such cases,  one expects steady values close
to  $\rho_0 \approx 3/5$ and $\phi_0 \approx 0.425$, respectively. The former 
value can be understood by recalling 
 the LDL phase has  density $\rho=3/4$ and  degeneracy 4 and coexists with the gas phase. 
Since their weights are equal ($1/5$) at the phase coexistence, the steady  $\rho_0 \approx 3/5$
is verified. The latter  value
is not easily understood, since the HDL phase is highly degenerated.
 Results for the tunneling between coexisting phases are shown for $L=12$. 
As in all previous cases, extreme frequencies give results deviating from  steady values, whereas
 the optimal frequency is signed for proper crossing between phases with
 accuracy improving as $R$ increases. However, in such case
each transition point study requires its own temperature schedule, since the
gas-LDL coexistence line is shorter than the LDL-HDL, ending at respective
distinct  critical temperatures $T_{c}=0.55$ and $0.825$,
respectively \cite{algs}. For each ${\mathcal T}_R$ we obtain the 
thermodynamic quantities, as shown
in parts \ref{fig7}$(c)$ and \ref{fig8} $(c)$. Note that
results are fully described by Eq. (\ref{eq1}) (continuous
lines). Also, both panels $(d)$ show 
the equivalence of results with those obtained
from the PT \cite{fioreprl}. However, larger replica numbers and
non adjacent replica exchanges were required for the system visiting
properly the phases when the PT is considered 
\cite{fioreprl}.

\section{Discussion and conclusion}

Simulated tempering with approximated weights, proposed by Park et. al. \cite{pande}, 
presents a great advantage  over other procedures, since the achievement of  
weights is readily obtained by executing simple and standard numerical simulations.
Aimed at unveiling and optimizing its performance,   we scrutinized
their main points focusing  in the regime  of first-order
transitions at low temperatures, in which
the existence of large trapping in metastable states makes the improvement 
of approximated weights a highly desired problem. Four  systems  having
available results from  parallel tempering, simulated tempering 
with free-energy weights, Wang-Landau method and cluster algorithms
have been considered  and thus they 
are important benchmarks for testing all obtained results.

In all cases,  results showed that  approximated weights improves the
efficiency of the ST as the replica number 
increases and  optimized sets  accelerating  the 
convergence toward the steady regime and promoting frequent tunneling between coexisting phases 
have been found.  We remark that the minimum value $R$ ensuring
proper sampling may depend on the system size and also the system temperature studied.
We believe that the  proposed recipe for the ST with approximated weights combined
with Eq. (\ref{eq1}) provides a very powerful method for dealing with discontinuous
transitions,  demanding relative short systems (as those considered here) and few control parameters 
\cite{fioreprl,fiorejcp2}. Although
not necessary (in conformity with the main ideas concerning Eq. (\ref{eq1})), the extension for larger
system sizes is straightforward. However, larger replica numbers or non-adjacent replica
exchanges can be required.
As a final comment, we remark that future method extensions include
polymeric models and off-lattice systems, in which tempering methods
have been extensively exploited, but the achievement of free-energy 
weights is difficult. These points should be addressed in future works.
\section{Acknowledgments} 

CEF acknowledges the financial support from CNPQ.
  


\begin{thebibliography}{99}  
\bibitem{binder} K. Binder and D. W. Heermann, Monte Carlo Simulation in
Statistical Physics (Springer-Verlag, New York Berlin Heidelberg, 1992).
           

 
\bibitem{metr} N. Metropolis, A. W. Rosenbluth, M. N.  Rosenbluth and
  A. H. Teller, J. Chem. Phys. {\bf 21}, 1087 (1953).

\bibitem{spinglass} K. Binder and W. Kob, {\it Glassy Materials and
    Disordered Solids: An Introduction to their Statistical Mechanics}
  (World Scientific, Singapoure, 2005).

\bibitem{proteins}  J. Skolnick and A. Kolinski,
  Comput. Sci. Eng. {\bf 3}(9/10), 40 (2001).

\bibitem {sw}R. H. Swendsen and J. S. Wang, Phys. Rev. Lett. {\bf 58},
  86 (1987), U. Wolff, Phys. Rev. Lett {\bf 62}, 361 (1989).


\bibitem{bouabci} M. Bouabci and C. E. I. Carneiro, 
Phys. Rev. B  {\bf 54}, 359 (1996).

\bibitem{berg} B. A. Berg and T. Neuhaus, Phys. Lett. B {\bf 267}, 249 (1991);  Phys. Rev. Lett. {\bf 68}, 9 (1992).
\bibitem{wang} F. Wang and D. P. Landau, Phys. Rev. Lett. {\bf 86}, 2050 (2001); 
 Phys. Rev. E {\bf 64}, 056101 (2001).

\bibitem{nemoto} K. Hukushima and K. Nemoto, J. Phys. Soc. Jpn. 
{\bf 65}, 1604 (1996).

\bibitem{parisi} E. Marinari and G. Parisi, Europhys. Lett. {\bf 19}(6), 451 (1992).

\bibitem{kone} A. Kone and D. A. Kofke, J. Chem. Phys {\bf 122},
206101 (2005).

\bibitem{helmut} H. G. Katzgraber, S. Trebst, D. A. Huse and
 M. Troyer, J. Stat. Mech. {\bf 3},  P031018  (2006).

\bibitem{sabo} D. Sabo, M. Meuwly, D. L. Freeman and J. D. Doll,
J. Chem. Phys {\bf 128}, 174109 (2008).
\bibitem{fiorejcp}
C. E. Fiore, J. Chem. Phys {\bf 135}, 114107 (2011).


\bibitem{juan} J. P. Neirotti, F. Calvo, D. L. Freeman and J. D. Doll, 
J. Chem. Phys. {\bf 112}, 10340 (2000).

\bibitem{juan1} F. Calvo, J. P. Neirotti, D. L. Freeman and
  J. D. Doll, J. Chem. Phys. {\bf 112}, 10350 (2000).

\bibitem{calvo} F. Calvo, J. Chem. Phys.
{\bf 123}, 124106 (2005).


\bibitem{fiore8}
C. E. Fiore, Phys. Rev. E  {\bf 78}, 041109 (2008).

\bibitem{fiore10}
C. E. Fiore and M. G. E. da Luz, Phys. Rev. E  {\bf 82}, 031104 (2010).

\bibitem{rosta} E. Rosta and G. Hummer, J. Chem. Phys. {\bf 131}, 165102 (2009); 
J. Chem. Phys. {\bf 132}, 034102 (2010).
\bibitem{maJCF}
C. Zhang and  J. P. Ma,
J. Chem. Phys, {\bf 129}, 134112 (2008).



\bibitem{ma}
C. Zhang and  J. P. Ma,
Phys. Rev. E {\bf 76}, 036708 (2007).



\bibitem{sauerwein}
R. A. Sauerwein and M. J. de Oliveira,
Phys. Rev. B, {\bf 52}, 3060 (1995).

\bibitem{fiore11}  
C. E. Fiore and M. G. E. da Luz, J. Chem. Phys {\bf 133}, 104904 (2010). 

\bibitem{pande}
S. Park and V. S. Pande, Phys. Rev. E {\bf 76}, 016703 (2007).


\bibitem{BEGMODEL}   M. Blume, V. J. Emery, and R. B. Griffiths,
  Phys. Rev. A {\bf 4}, 1071 (1971),  W. Hoston and
  A. N. Berker, Phys. Rev. Lett. {\bf 67}, 1027 (1991).
\bibitem{bell}
G. M. Bell and D. A. Lavis, J. Phys. A {\bf 3}, 568 (1970).

\bibitem{fiore-m}
C. E. Fiore, M. M. Szortyka, M. C. Barbosa and V. B. Henriques,
J. Chem. Phys {\bf 131}, 164506 (2009).


\bibitem{algs} 
A. L. Balladares, V. B. Henriques, and M. C. Barbosa, 
J. Phys. C {\bf 19}, 116105 (2007). 

\bibitem{alg} 
V. B. Henriques and M. C. Barbosa, 
Phys. Rev. E {\bf 71}, 031504 (2005). 
\bibitem{cluster2}
C. E. Fiore and C. E. I. Carneiro, Phys. Rev. E {\bf 76}, 021118 (2007).



\bibitem{claudio} C. J. Silva, A. A. Caparica and J. A. Plascak,
Phys. Rev. E  {\bf 73}, 036702 (2006).


\bibitem{alexandra} 
A. Valentim, M. G. E. da Luz and C. E. Fiore, 
Comp. Phys. Comm. {\bf 128}, 2046 (2014).







\bibitem{fiore-m2}
M. M. Szortyka, C. E. Fiore, V. B. Henriques and M. C. Barbosa,
J. Chem. Phys {\bf 133}, 104904 (2010).


\bibitem{fioreprl}
C. E. Fiore, M. G. E. da Luz,
Phys. Rev. Lett. 107  230601, (2011).

\bibitem{fiorejcp2}
C. E. Fiore, M. G. E. da Luz, J. Chem. Phys. 138  014105, (2013).


\bibitem{ferdinand}
B. Kaufman, Phys. Rev. {\bf 76}, 1232 (1949); A. E. Ferdinand and M. E. Fisher, 
Phys. Rev. {\bf 185}, 832 (1969).







\end{thebibliography}
\end{document}